\documentclass[aps,prb,showpacs,twocolumn]{revtex4}

\usepackage{amsmath}
\usepackage{epsfig}
\usepackage{rotating}

\begin{document}

\title{Vibrational properties of single-wall nanotubes and
       monolayers of hexagonal BN}

\author{D. S\'{a}nchez-Portal~\cite{byline}}
\affiliation{Centro Mixto CSIC-UPV/EHU and 
Donostia International Physics Center (DIPC) \\
Paseo Manuel de Lardizabal 4, 20018 Donostia-San Sebastian, Spain}
\author{E. Hern\'{a}ndez~\cite{ehe}}
\affiliation{Institut de Ci\`{e}ncia de Materials 
de Barcelona, ICMAB - CSIC, \\
Campus de Bellaterra, 08193 Bellaterra, Barcelona,
Spain}

\date{\today}

\begin{abstract}
We report a detailed study of the vibrational properties of BN single-walled
nanotubes and of the BN monolayer. Our results have been obtained from a
well-established Tight-Binding model complemented with an electrostatic 
model to account for the long-range interactions arising from the polar nature 
of the material, and which are not included in the Tight-Binding model.
Our study provides a wealth of data for the BN monolayer and nanotubes, such as
phonon band structure, vibrational density of states, elastic constants, etc. 
For the nanotubes we obtain the behavior of the optically active modes as a 
function of the structural parameters, and we compare their frequencies 
with those derived from a zone-folding treatment applied to the phonon
frequencies of the BN monolayer, finding general good agreement between the
two.
\end{abstract}

\pacs{63.22.+m, 62.25.+g, 78.30.Na, 78.20.Bh}

\maketitle

\section{Introduction}

The discovery of the fullerenes~\cite{kroto} in the mid 1980's, and 
especially that of carbon nanotubes~\cite{iijima} in the early 1990's
has fueled a frenzy of research activity in the field of carbon 
nanostructures, doubly motivated by their large potential for practical
applications on one hand, and because these systems constitute a new 
open field for basic research at the nanoscale on the other. Nanotubes
can be single-walled or multi-walled, and are quasi-1D structures having
large aspect ratios (the quotient of length over diameter). Single-wall carbon
nanotubes (SWNTs) can be either semiconducting or metallic, depending on their
structural parameters, and can be chiral or achiral. They have large
thermal conductivity, and have the highest Young's modulus ever 
measured~\cite{ebbesen1,lieber,ebbesen2,forro}.
Nanotubes can be filled with other elements, compounds or
even fullerenes in their internal channel, forming structures similar to 
nanoscale coaxial cables. All these properties confer nanotubes a large
potential for technological applications, some of which have already
been demonstrated. Nanotubes have been used to fabricate field 
emission devices~\cite{field}, tips for scanning probe microscopy 
instruments~\cite{tips}, constituents of nanoelectronic devices~\cite{dekker},
gas sensors~\cite{sensors}, composite reinforcement~\cite{composites},
lubrication, etc. It is therefore not surprising that these systems 
have been and continue to be the focus of considerable interest; the
large number of review articles~\cite{reviews} and monographs~\cite{books}
devoted to the different aspects of fullerenes and nanotubes in recent years 
provides further proof of this interest.

One of the probes most frequently used to characterize carbon nanotube 
samples is vibrational spectroscopy, and specifically Raman 
spectroscopy~\cite{dresselhaus:eklund}. Raman spectroscopy offers 
experimentalists a rapid way of estimating the diameter distribution
of tubes present in a sample, because the low frequency radial breathing
mode (RBM) of SWNTs has a frequency which is inversely proportional to
the square of the nanotube diameter, independently of the nanotube structure,
determined by the nanotube indices (n,m). Also, the high frequency G~band
modes resulting from atomic vibrations in the nanotube shell, can help to
distinguish between metallic and semiconducting nanotubes in the 
sample~\cite{Gband}.
Jorio and coworkers~\cite{jorio} have recently demonstrated the ability
of Raman spectroscopy to provide full structural determinations of isolated
carbon SWNT's.
The diverse aspects of phonons in carbon SWNT's and the use of Raman
spectroscopy as a tool for their characterization have been recently
reviewed by Dresselhaus and Eklund~\cite{dresselhaus:eklund}.

Soon after the discovery of carbon nanotubes it was speculated that other 
materials could possibly form similar nanostructures, since there are several
elements and many compounds which form layered structures bearing some 
resemblance to graphite. The most obvious candidate was hexagonal BN (h-BN), 
which was predicted on the basis of theoretical 
calculations~\cite{corkill,blase}
to be capable of forming nanotubes, a prediction which was later 
corroborated experimentally by the synthesis of such
nanotubes~\cite{bn_synthesis}. Today we know that many other structures
can form nanotubes ($\mbox{MoS}_2$, $\mbox{WS}_2$~\cite{MoS2}, Bi~\cite{Bi},
\ldots); nevertheless it is still the case that carbon nanotubes continue to 
attract a larger interest, but certainly these other structures are 
interesting in their own right, and may be able to offer different 
possibilities for technological applications that carbon nanotubes cannot
provide. Both multi-walled and single-walled~\cite{loiseau} BN nanotubes can
now be readily synthesized, and these tubes are uniformly insulating, tending 
to have a zig-zag structure.

As other types of nanotubes become more common, many of the 
characterization tools and techniques extensively used in the case of 
carbon nanotubes will undoubtedly also find application in the study
of these other structures. In particular, vibrational spectroscopy, which
has proved to be such a useful tool in the case of carbon nanotubes, is
likely to prove useful also in these cases. In this context, it is interesting
to consider the phonon properties of BN nanotubes from a theoretical
point of view; their study will help to develop the characterization
potential of spectroscopic techniques when they are applied to BN
nanotubes. The aim of this paper is to provide such a theoretical study.
We report extensive theoretical calculations of the vibrational properties
of a h-BN layer and of a number of BN nanotubes having diameters in
the range of 0.4 to 2~nm, including zig-zag, arm-chair and several chiral
nanotubes. From our results for the flat sheet we have performed a 
zone-folding analysis in order to predict the vibrational properties of
the tubular structures from those of the flat layer, a technique which has
been frequently used in the case of carbon nanotubes. The direct calculation
of the vibrational properties of tubular structures allow us to compare the
predictions of zone-folding with the actual results, and thus gauge the
applicability of the zone-folding approach for BN nanotubes.
Other authors have previously considered the vibrational~\cite{miyamoto}
and elastic~\cite{kudin} properties of isolated h-BN monolayers and of
bulk h-BN~\cite{kresse,dynamic-charges}. 
Some aspects of the elasticity of BN nanotubes have also been studied in 
references~\onlinecite{elastic_tubes} and~\onlinecite{kudin}. 
However, the study of 
the vibrational properties in the 
case of BN nanotubes has not, to our knowledge, been addressed to date.

The structure of the paper is as follows. In sec.~\ref{sec:model} we 
describe the model and calculation procedure used in our study; later
in sec.~\ref{sec:results} we discuss our findings, starting with a 
thorough description of the vibrational properties of an isolated
h-BN monolayer, followed by the results obtained for the nanotubes,
establishing a comparison between the results obtained from the zone-folding
analysis and those obtained from direct calculation. Our summary and
conclusions are discussed in sec.~\ref{sec:concs}.

\section{Model and computational procedure}
\label{sec:model}

\begin{figure}[]
\begin{minipage}[t]{8.0cm}
\begin{center}
\leavevmode
\epsfxsize=8.5cm
\begin{turn}{-90}
\epsffile{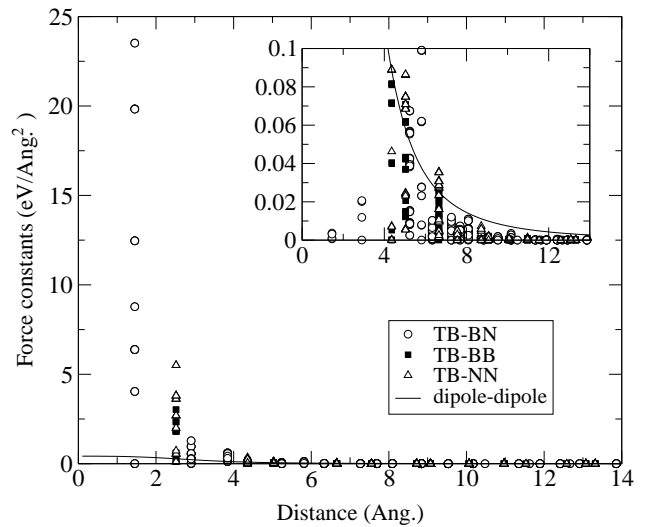}
\end{turn}
\end{center}
\end{minipage}
\caption{Force constants derived from the TB model (symbols) and the typical
behavior of the electrostatic
interactions (approximated in the plot by $f(r)\;Z_{\parallel}^2/(\epsilon \:r^3)$)
(solid line),
as a function of distance. The inset shows more
clearly the behavior for large distances. }
\label{fig:constants_vs_distance}
\end{figure}

All calculations have been performed using the non-orthogonal
Tight-Binding (TB)~\cite{tb_review} parametrization of Widani
{\em et al.\/}~\cite{widany}. This is a parametrization following the
scheme proposed by Porezag and coworkers~\cite{porezag}. This model
employs a basis set consisting of one~S and three~P functions per
atom, to represent valence states, and the resulting Hamiltonian
and overlap matrix elements extend up to a range of approximately
5~\AA. The core-core repulsion is modeled by means of a simple
pair-repulsive potential. The original work of Porezag {\em et
al.\/}~\cite{porezag} described a non self-consistent TB model;
although an extension of the methodology has been proposed which
incorporates some degree of self-consistency~\cite{frauenheim},
here we only have used the model of Widany~\cite{widany} in
its non self-consistent form. It is worth commenting that in spite of
its simplicity this model has proved to be quite
successful~\cite{widany,elastic_tubes}.

The equilibrium structures of the tubes were obtained by careful
minimization of the total energy with respect to both the atomic
coordinates and the lattice constant along the tube axis.
For the phonon calculations, we then computed the force-constant
matrix in real space using a finite-difference approach~\cite{finitedifferences}.
We used atomic displacements of 0.02~\AA, and the force constants were taken
as the average of the results obtained with positive and negative displacements,
in order to eliminate anharmonic effects. Since all nitrogen and boron atoms
are equivalent by symmetry in the tubes,  
we only calculated the force constants for one nitrogen and one boron 
atom in the supercell and generated the rest of the matrix using the
symmetry operations.

The force-constants have to be calculated between a given atom and all
the other atoms in the system. However, since the interactions 
in our TB model decrease rapidly with distance, only the elements
with atoms sufficiently close have to be computed. To do so, we set up 
a supercell large enough that a sufficient number of neighbors 
is included. It must be kept in mind that, in the supercell geometry, 
a given atomic displacement in the central cell is always accompanied 
by the same displacement of all image atoms. We need, therefore, 
to use a supercell such that the effect of the image displacements
is negligible. For this purpose we have used supercells containing
six unitcells in the case of (n,n) and (n,0) tubes, and 
two and one unitcells, respectively, 
in the case of the (10,5) and (10,7) tubes. The supercells contain hence
a minimum of 144 atoms for the (6,0) and (6,6) tubes, and a maximum 
of 384 atoms in the case of the (16,0) tube.

The above described TB model does not by itself incorporate any
long-range electrostatic interactions. Such interactions are
important when considering the vibrational properties of
polar materials, because they will make a long-range contribution
to the force constants. Therefore we have opted for correcting
this shortcoming in the TB model in a physically sound but
otherwise empirical way.
The key observation in order to introduce the effects of the
electrostatic interactions is that,
when an atom $i$ suffers a displacement ${\bf u}(i)$
from its
equilibrium position,
a net electric dipole of magnitude
${\bf p}^i= \sum_\mu Z^i_{\nu \mu} u^i_\mu$ appears associated
with this atomic movement~\cite{Gonze}.
Here $Z^i_{\nu \mu}$
is the Born effective charge tensor~\cite{Gonze} of the atom $i$.
As a consequence,
if two atoms $i$ and $j$ are simultaneously displaced from their positions,
besides the interaction energy given
by our TB model, it is necessary to include the long-range interaction between
the two electric dipoles generated. This gives rise to a new term in
the force-constants matrix of the form~\cite{Gonze}
\begin{eqnarray}
C^{LR}_{\nu \mu}({\bf r}_{ij})= & \\  \nonumber
f(r_{ij}) & \sum_{\nu^\prime \mu^\prime}
\frac{Z^i_{\nu \nu^{\prime}} Z^j_{\mu \mu^{\prime}} }
{\epsilon } \left( \frac{\delta_{\nu^\prime \mu^\prime}}{r_{ij}^3}
-3 \frac {r_{ij \nu^\prime} r_{ij \mu^\prime}}{r_{ij}^5} \right),
\label{eq:longrange}
\end{eqnarray}
where the superindex {\em LR\/} in $C^{LR}_{\nu \mu}({\bf r}_{ij})$ indicates 
that this is the long-range contribution to the force constants;
$f(r_{ij})$ is a switching function (see below),
$\epsilon$ is the dielectric
constant, ${\bf r}_{ij}$ is the vector going from atom $i$ to atom $j$,
and $r_{ij}$ is the distance between these two atoms.
The
switching function is designed to come into play at distances sufficiently
large so as to not affect the TB model; it takes the form
\begin{eqnarray}
f(r) = 1 - e^{-(r/r_c)^3}.
\label{eq:switching}
\end{eqnarray}
Following the usual approach~\cite{Gonze}, it is also necessary to modify the
on-site elements of
the force matrix
in order to satisfy the acoustic sum rule. Therefore, we take
\begin{eqnarray}
\label{eq:onsite}
C^{LR}_{\nu \mu}({\bf 0}_{ii})=-\sum_{j \neq i} C^{LR}_{\nu \mu}({\bf r}_{ij}).
\end{eqnarray}
The Born effective charge tensor has been taken from recent {\it ab initio}
density functional calculations~\cite{dynamic-charges}: $Z^B_{\perp}$=
$-Z^N_{\perp}$=0.82 for out-of-plane, and  $Z^B_{\parallel}$=
$-Z^N_{\parallel}$=2.71 for in-plane displacements~\cite{tensorfortubes}.
We have taken $\epsilon = 4$ and $r_c = 2.6$~\AA, the distance of
separation between second nearest neighbors in the h-BN
layer. We have not attempted to fit these values in order to
improve agreement of our results with experimental measurements.
Rather, our approach has been to take reasonable values for them
in order to estimate the effect of the polar interactions
on the vibrational properties of the BN monolayer and in the
nanotubes. Our chosen value for $\epsilon$ is close to the
experimental one for h-BN (4-5 depending on the
direction~\cite{geick,dynamic-charges}),
and the value of $r_c$ used in Eq.~(\ref{eq:switching}) is motivated by
the fact that force constants arising from the TB model
alone decay very rapidly with distance, and they are already
very small at this value of $r_c$, as can be seen in
Fig.~\ref{fig:constants_vs_distance}. By choosing $r_c$ equal
to the second nearest neighbor distance we ensure a smooth
switching between the short-range TB model and the long-range
electrostatic one, with minimum interference between them.

Once we have the force-constant matrix in real space, 
we can calculate the dynamical matrix in reciprocal space, and 
diagonalize it to obtain the phonon modes and frequencies. 
In the case of bulk 
polar systems the computation of the dynamical matrix in reciprocal space has to be 
made with some care. The reason is that, due to the long-range 
nature of the dipole-dipole interaction in Eq.(\ref{eq:longrange}),
a different limit for the dynamical matrix is obtained,
as the phonon wavevector ${\bf k}$ approaches ${\bf \Gamma}$, 
in the cases of longitudinal and transversal vibrations.
In our case however, due to the reduced dimensionality 
of the systems considered (a monolayer and single-walled tubes), 
the Fourier transform of the dynamical matrix 
can be performed without further complications by simply 
adding up the elements of the real-space matrix
with the appropriate phase factors. Given its simple analytical form,  
we can include a large number of neighbors (we tipically take all neighbors
within a radius of $\sim$200~\AA) in the summation involving 
the dipole-dipole interactions and guarantee a good convergence. 
The same limit is found for both 
polarizations when ${\bf k} \rightarrow {\bf \Gamma}$, and there are no splittings
between in-plane 
longitudinal and transversal optical modes at ${\bf \Gamma}$
(see Appendix~\ref{nosplitting}).  
 
\section{Results and discussion}
\label{sec:results}

\begin{figure*}[!]
\epsfxsize=7.0cm
\epsffile{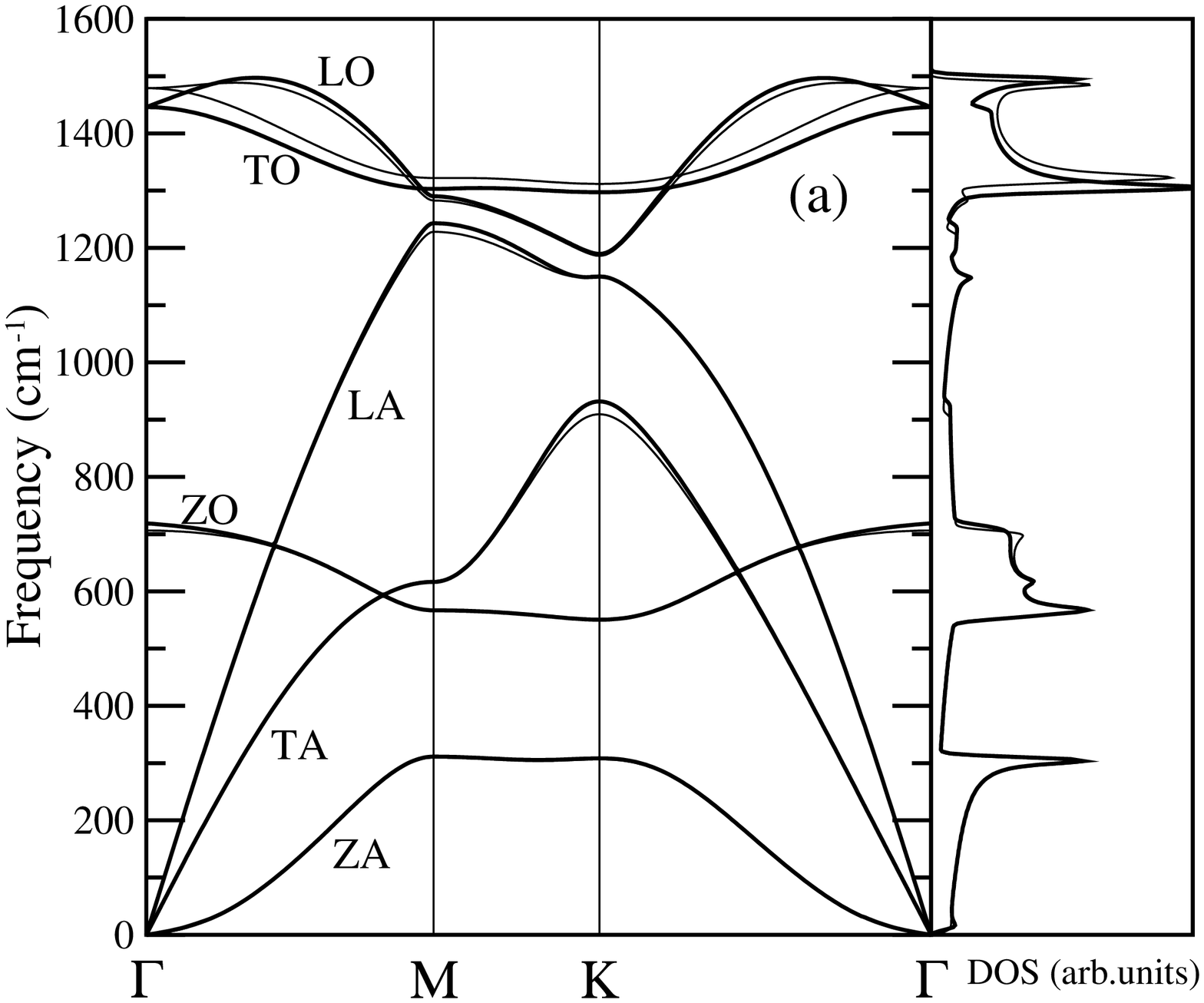}
\epsfxsize=8.0 cm
\epsffile{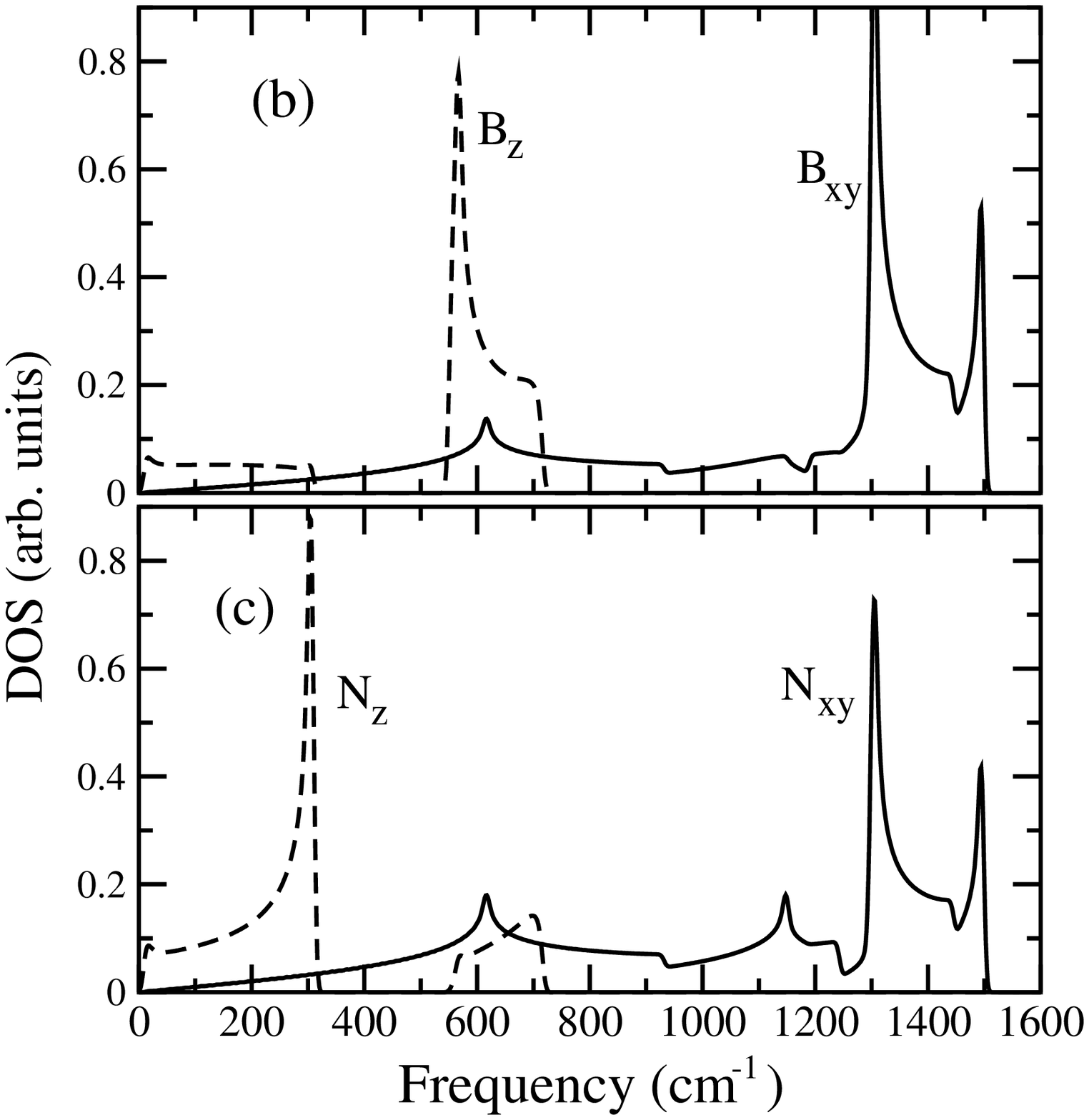}
\caption{(a) Phonon band structure of the BN hexagonal monolayer (right) and
the corresponding density of vibrational states (left); thick lines correspond
to band structure and density of states calculated with electrostatic
interactions included, while thin lines give results obtained only with
the TB model. (b) Projected density of states on the B atoms; continuous
line represents the density of states projected onto the plane of the monolayer,
while the discontinuous line gives the density of states projected
perpendicular to the plane of the monolayer. (c) As in (b) for the N atoms.}
\label{fig:sheet}
\end{figure*}

One of the aims of this work is to judge the validity of the zone-folding
approach for obtaining phonon information of tubular structures from the 
phonon band structure of the 
flat sheet, a procedure which has 
been frequently used in the case of carbon nanotubes~\cite{dresselhaus:eklund}. 
To this end we have 
performed a phonon analysis of the flat sheet, and of several nanotubes
(arm-chair, zig-zag and chiral) so as to compare the zone-folding results
with those obtained directly from the tubular structures. In the following
we will first present the results obtained for the flat sheet, followed
by those for a selected set of nanotubes. The comparison between these 
latter results and the zone-folding predictions will be then presented.

\subsection{Flat sheet}
\label{sub:sheet}

Using the TB model discussed in section~\ref{sec:model}, with and without
the dipolar interactions, with a supercell
containing 192 atoms we have obtained the phonon dispersion curves
and vibrational density of states for a flat sheet of h-BN,
plotted in Fig.~\ref{fig:sheet}. In order to check the convergence
of the results with respect to the supercell size, we performed a test with 
a supercell containing 320 atoms; the results obtained with the smaller
cell are indistinguishable from those of the larger cell, and hence all 
subsequent analysis was performed with the 192 atom cell results. We also
performed a more detailed analysis of the convergence of the phonon bands
with respect to the lower threshold of the elements of the real-space
force-constant matrix included in the construction 
of the dynamical
matrix. 
We focus first on the convergence with respect to the TB force constants. 
We found that including all the couplings 
coming from the TB hamiltonian 
with absolute value larger
than $1.2 \times 10^{-2}\ \mbox{eV/\AA}^2$ was 
necessary to adequately converge 
all bands. This threshold corresponds to a spatial cutoff of 6.7~\AA, 
which includes all atoms within the tenth nearest-neighbor shell. This
large cutoff is required to correctly 
reproduce the quadratic behavior of the 
lowest acoustic band [labeled ZA in Fig.\ref{fig:sheet}] in the limit 
${\bf k} \rightarrow {\bf \Gamma}$,
with a zero sound velocity (slope of the band) 
as {\bf k} approaches ${\bf \Gamma}$. 
This quadratic behavior of the ZA band
is a consequence of the
fact that, at least to the lowest order in its amplitude, 
the
strain energy created by this vibration is solely associated with  
the  curvature that this 
out-of-plane bending mode induces in the layer. Optical bands typically
converge much faster ({\em ca.\/} including interactions with neighbors
within the third or fourth nearest-neighbor shell). It is worth mentioning
in this respect that in the earlier theoretical work of Miyamoto
{\em et al.\/}~\cite{miyamoto} a supercell which only permitted the 
inclusion of second nearest neighbor interactions was used due to 
computational limitations (Miyamoto and coworkers used first-principles
methods, which are significantly more costly, to evaluate the force-constant 
matrix). Although inclusion of up to second nearest neighbor interactions is
sufficient to adequately converge most bands, it fails to reproduce the 
over-bending (see below) of the longitudinal-optical (LO) band, for which the 
maximum occurs not at ${\bf \Gamma}$ but approximately 
half way along the symmetry lines ${\bf \Gamma-M}$ and ${\bf \Gamma-K}$, and 
is insufficient to reproduce the quadratic behavior of lowest acoustic 
band around ${\bf \Gamma}$.

Let us now comment briefly on the convergence with respect to the long-range
electrostatic interactions.
As already commented in the previous section, 
due to its simple analytical form we have used 
a large cut-off of 200~\AA\ for the dipole-dipole 
interactions. However, we have 
checked that already with a cut-off of $\sim$25~\AA\ the results are 
well converged. Further reducing the  
interaction range starts to have an appreciable effect. Not only the 
position of the optical branches at and close to ${\bf \Gamma}$ start
to shift from their converged values, but
the finite slope at ${\bf \Gamma}$ that the LO mode must exhibit
as a consequence of the long-range nature of the interactions also disappears
(see Appendix~\ref{nosplitting}). 
  
As can be appreciated in Fig.~\ref{fig:sheet}, due to the 
two-dimensional character of the system, there is a separation between
in-plane and out-of-plane modes, the latter being the lower frequency
ones, with a maximum at ${\bf \Gamma}$ of 719~$\mbox{cm}^{-1}$ with 
dipolar interactions included (706~$\mbox{cm}^{-1}$ if these are not 
included). This frequency
separation is to be expected, since in-plane modes excite both bond-stretchings
(hard) and bond-bendings (soft), while out-of-plane modes result mostly in
bond-bending motion and very little bond-stretching.
The highest frequency bands, the LO and the transversal-optical (TO), 
have a frequency at 
${\bf \Gamma}$ of 1446~$\mbox{cm}^{-1}$ when the dipolar interactions are
included (1479~$\mbox{cm}^{-1}$ without). The 
maximum of the LO band occurs not
at ${\bf \Gamma}$, but at an intermediate point away from the Brillouin
zone edges, as pointed out above. This is known as over-bending, and the 
presence of over-bending in the flat sheet phonon band structure can have 
interesting consequences for the phonon band structure of the
nanotubes, as it may lead to the appearance of new modes at ${\bf \Gamma}$ 
having higher frequencies in the tubular structures than 
in the flat sheet~\cite{kasuya}.
Note that including the dipolar interactions increases the over-bending,
which is much smaller when the electrostatics are not appropriately
accounted for.
Another noteworthy characteristic of the phonon band structure 
displayed in Fig.~\ref{fig:sheet}(a) is the splitting of approximately
250~$\mbox{cm}^{-1}$ existing between the ZO and ZA bands along
{\bf M-K}. In a homopolar hexagonal sheet such as graphene these two bands
cross at {\bf K}~\cite{sanchez-portal}.

\begin{table}[!]
\begin{center}
\begin{tabular}{cl}
\hline \hline
$C_{xxxx} \equiv C_{11}$ & 0.343 (0.322)  \\
$C_{xxyy} \equiv C_{12}$ & 0.119   \\
$C_{xyxy} \equiv C_{66}$ & 0.107 (0.123)  \\
$Y$  & 0.302  \\
$\sigma$ & 0.347 \\
\hline \hline
\end{tabular}
\end{center}
\caption{Elastic constants and Young modulus $Y$, in units of TPa~nm,
and  Poisson ratio $\sigma$
of the flat BN hexagonal sheet, as
calculated directly from the force-constant matrix given by our model.
The results obtained
from the sound velocities of the TA and LA bands (see Eq.(\ref{eq:velocities}))
are shown in brackets.}
\label{table:constants}
\end{table}

From the slopes of the transversal-acoustic (TA) and longitudinal-acoustic (LA) 
bands in the limit 
${\bf k} \rightarrow {\bf \Gamma}$ we deduce a sound velocity of 
13~Km/s and 21~Km/s, respectively. As already pointed out
above, the ZA band exhibits a $k^2$ behavior in this limit, as expected,
and therefore the sound velocity associated to this band is zero. 
Fitting the frequencies below 250~$\mbox{cm}^{-1}$ to the expression
$\omega$=$2 \pi \nu$=$\delta k^2$+$\gamma k^3$, we can estimate
a value of $\delta \approx$ 1820 $\pm$ 60~$\mbox{cm}^{-1}$\AA$^2$=
5.5$\pm$0.2 $\times$10$^{-7}\mbox{m}^2\mbox{s}^{-1}$. This parameter
is interesting because it allows to estimate the energy necessary 
to roll up the BN sheet in order to form the nanotubes. 
Simple arguments show that the strain energy per atom can
be approximated by E$_{st}$=C/$r^2$ with C=$\delta^2(m_N+m_B)$/4,
being $m_N$ and $m_B$ the masses of the nitrogen and
boron atoms, respectively, and $r$ the tube radius. 
The value obtained for 
C in this way is 1.91~eV\AA$^2$, which 
is in reasonable agreement with the magnitude obtained from
total energy calculations performed using both,
a tight-binding hamiltonian
similar to the one utilized here~\cite{elastic_tubes}, and first-principles density
functional calculations~\cite{blase,kudin}. 
A similar estimation in the case
of graphene~\cite{sanchez-portal} ($\delta \approx$ 6 
$\times$10$^{-7}\mbox{m}^2\mbox{s}^{-1}$) leads to a value
C$\approx$ 2.3~eV\AA$^2$. Therefore, already
looking at the phonon band structures it is possible to
find an indication that the strain energy 
is smaller in the case of BN tubes than for carbon nanotubes, 
a result which is confirmed by more sophisticated 
calculations.~\cite{blase,elastic_tubes,kudin,largetubes}

Using the calculated force-constants matrix
we can directly obtain the elastic constants~\cite{ashcroft:mermin} of the
flat sheet, which are given in Table~\ref{table:constants}. In bulk 
three-dimensional systems it is customary to quote elastic constants in
units of pressure, due to the inverse equilibrium cell volume factor 
that appears in the definition of the elastic 
constants~\cite{kittel,ashcroft:mermin}. However, in two-dimensional 
systems the
definition of the cell volume is arbitrary, and it is therefore more
appropriate to use cell area in the definition of the elastic 
constants~\cite{elastic_tubes}; hence, we provide our results in 
units of pressure times length. 
Comparing with previous calculations in the literature,
our result for C$_{xxxx}$ is in reasonable agreement 
with the value of 
0.271~TPa~nm obtained for the in-plane stiffness~\cite{stiffness} by
Kudin and coworkers using density-functional
theory and Gaussian type orbitals~\cite{kudin}.
Our result is also in quite good agreement with the value
of 0.309~TPa~nm that can be deduced 
from the plane-wave density-functional calculation of 
the elastic constants of bulk h-BN
reported by Ohba {\it et al.}
in Ref.~\onlinecite{dynamic-charges} 
(we use here the calculated interlayer distance of 3.25~\AA\ 
to translate the data from the bulk 
to the monolayer geometry). 
However, the comparison is 
poorer for the C$_{xyxy}$ elastic constant, 
for which the data of Ohba and coworkers 
translate into 0.055~TPa~nm for the monolayer (roughly half our value). 
The origin of this large discrepancy is unclear, 
although it
might be related to the interactions with neighboring layers, which are not
present in our case.

For a hexagonal two-dimensional crystal only
two elastic constants are required~\cite{landau:lifshitz}, and due to the
underlying crystal symmetry we obtain the following rule:
\begin{eqnarray}
C_{xxxx} = C_{xxyy} + 2 C_{xyxy}.
\label{eq:check}
\end{eqnarray}
This provides an internal consistency check for our results, which, as can be
seen from the results in  Table~\ref{table:constants}, is satisfactorily
obeyed [the numerical value obtained for $C_{xxxx}$ using $C_{xxyy}$ and
$C_{xyxy}$ in Eq.~(\ref{eq:check}) is within 3~\% of its calculated 
value, given in Table~\ref{table:constants}]. A second consistency check 
is provided by the comparison of these values of the elastic constants
with those obtained from the sound velocities reported above
using the identities
\begin{eqnarray}
v^{2D}_{LA}=\sqrt{ {C_{xxxx} \over \rho}}, \;
v^{2D}_{TA}=\sqrt{ {C_{xyxy} \over \rho}}, 
\label{eq:velocities}
\end{eqnarray}
where $\rho$ is the surface mass density of the sheet.
These data are
also given in Table~\ref{table:constants}. As can be seen there, the
values obtained in these different fashions are in rather good agreement.
From the elastic constants we 
can also calculate~\cite{landau:lifshitz} the Poisson ratio,
\begin{eqnarray}
\sigma = \frac{C_{xxyy}}{C_{xxxx}},
\label{eq:poisson}
\end{eqnarray}
and Young's modulus,
\begin{eqnarray}
Y = \frac{C_{xxxx}^2 - C_{xxyy}^2}{C_{xxxx}}.
\label{eq:young}
\end{eqnarray}
The values obtained for $\sigma$ and $Y$ are also given in 
Table~\ref{table:constants}. These values can be compared with those 
obtained previously for BN nanotubes using the same TB model~\cite{elastic_tubes}; 
the value 
of $\sigma$ obtained here is a little bit larger than that obtained for
the nanotubes (0.263), but the agreement with the tube Young's modulus 
(0.298-0.310 TPa nm depending on the tube diameter) is nearly perfect. 
We also point out that the inclusion of the dipolar interactions does not
affect the values of the sound velocities or elastic constants, as expected
from the fact that they only influence the high frequency optical phonon bands,
but not the acoustic ones, as can be seen in Fig.~\ref{fig:sheet}.

At the zone edge the ZA band corresponds mostly to N atom out-of-plane
displacements, while the ZO mode corresponds to B displacements. This is 
most clearly seen in Fig.~\ref{fig:sheet}(b) and (c), which shows the partial
vibrational density of states (DOS) separated into its in-plane and
out-of-plane atomic contributions.

The frequency of the Raman and infrared-active (IR) modes in 
bulk h-BN has been 
investigated by several authors~\cite{geick,kuzuba,nemanich,hoffman} 
obtaining similar results. 
The values reported for the high energy 
Raman-active E$_{2g}$ mode are 
1366~\cite{nemanich,kuzuba}, 1367~\cite{hoffman}, 
and 1370~\cite{geick}~$\mbox{cm}^{-1}$. 
For the IR modes the values given in the literature are: 1367~\cite{geick}
and 1383$\pm$5~\cite{hoffman}~$\mbox{cm}^{-1}$ for the E$_{1u}$(TO) mode,
1610~\cite{geick}~$\mbox{cm}^{-1}$ for the E$_{1u}$(LO) mode, and
770$\pm$3~\cite{hoffman} and 783~\cite{geick}~$\mbox{cm}^{-1}$
for the
A$_{2u}$(TO) mode. At this point it is important to note that, since   
our calculations refer to a free-standing (isolated)
flat sheet of h-BN rather than to the bulk material, the
comparison with these experimental observations has to be made with
some care. In analogy to the case of graphite, 
and assuming relatively weak 
interlayer interactions, the frequencies of a single BN 
sheet should be closely 
related to those of the bulk. However, there is an important 
exception related
to the role of the long-range electrostatic interactions. In a strictly 
two-dimensional (2D) system 
like the one treated here ({\it i.e.} a sheet of monatomic thickness), 
the electrostatic interactions do not result in a macroscopic term at 
${\bf \Gamma}$ for in-plane modes, and hence we do 
not obtain any LO-TO splitting in 
that limit (see Appendix~\ref{nosplitting}).
This striking difference is quite notorious, for example, comparing our 
phonon band structure
in Fig.~\ref{fig:sheet}(a) with the calculations for bulk h-BN diplayed
in Fig.~3 of Ref.~\onlinecite{kresse}. 
The absence of LO-TO splitting for the systems of 
reduced dimensionality considered here
turns out to be quite important 
when considering the applicability of the zone-folding approach to the 
vibrational properties of BN nanotubes, a point to which we will return below.

Our results for the planar BN sheet 
have, therefore, to be compared with the TO frequencies
of the bulk. We obtain for the highest modes at ${\bf \Gamma}$ 
a frequency of 1446~$\mbox{cm}^{-1}$, to be compared against the 
empirical values of the E$_{2g}$ and E$_{1u}$(TO) modes,  
and 719~$\mbox{cm}^{-1}$ for the ZO mode, to compare with the measurements 
for the A$_{2u}$(TO).
The discrepancies are therefore
smaller than 10~\%, which we regard as acceptable, given the simplicity
of the TB model employed in this work. It is noteworthy that the effect of the
inclusion of the long-range electrostatic interactions improves the 
comparison with experiment, correcting the
under-estimation of the ZO($\bf \Gamma$) frequency and the over-estimation
of the TO($\bf \Gamma$) and LO($\bf \Gamma$) value, as well as giving a more
realistic over-bending of these modes along ${\bf \Gamma}-{\bf M}$ and
${\bf K}-{\bf \Gamma}$. We should emphasize here that no
empirical information on vibrational or structural properties has
been used in the construction and parametrization of the model~\cite{widany}.
As pointed out above, it is also important to take into account that
at least some of the discrepancy
may not be attributed to the model, but to the geometry used in the
supercell calculation. The presence of near-by sheets in the experiment 
is likely to soften somewhat
the in-plane LO and TO modes, while hardening the ZO mode, which could 
account for some of the discrepancy between our results and the experimental
ones. 
Discrepancies of similar size also 
occur between
the first-principles values of Miyamoto {\em et al.\/}~\cite{miyamoto}
and the experimental values. The overall topology of our phonon band
structure is in good agreement with the results of Miyamoto {\em et al.\/},
although we cannot make a direct comparison due to the unusual shape of
the h-BN supercell used in that work.
Kern {\em et al.\/}~\cite{kresse} have also performed first-principles density 
functional calculations
of the phonon properties of both cubic and h-BN. Our phonon band
structure is in reasonable agreement with that obtained by them, bearing
in mind the different methodologies and geometries used in their
work and ours.

Rokuta {\em et al.\/}~\cite{rokuta} have obtained the phonon spectra of a 
BN monolayer deposited over a series of metal surfaces (Ni, Pd and Pt) using
high resolution electron energy loss spectroscopy. Again, there is overall
good agreement between the results reported by Rokuta and coworkers, and
our own. Interestingly, they find no splitting at ${\bf \Gamma}$ between the
LO and TO modes when the BN monolayer is placed over a Ni(111) surface, 
but some splitting is seen when the other metals are used. 
This result was considered as quite surprising by these authors, especially
taking into account that Ni(111)
was the sole substrate on which BN formed well ordered commensurate monolayers, 
and it was suggested 
that this is because Ni can effectively screen the polarization field of 
the LO mode, while Pd and Pt cannot. However, we maintain that the polarization
field only arises in 3D systems; in the 2D case the electrostatic 
interactions do not give rise to a macroscopic field for in-plane modes, 
{\em i.e.\/} the
field has zero component at ${\bf \Gamma}$, and hence there is no splitting
of the TO and LO bands in that limit. However, the electrostatic interactions
do give rise to non-zero components of the field in regions of the Brillouin
zone away from ${\bf \Gamma}$, and these result in a higher over-bending
when the electrostatic interactions are included. Another consequence of
the inclusion of the electrostatic interactions is the fact that the LO
mode approaches ${\bf \Gamma}$ with a finite slope, which does not occur
when the electrostatics are not included [see Fig.~\ref{fig:sheet}(a)].

Prompted by this experimental study, we have considered the effect of the 
presence of an ideal metal surface in the proximity of the BN layer on the
phonon properties of the latter. The metal surface is simply considered as
a medium which generates dipolar images of the BN monolayer. The net result
is to effectively shorten the range of the electrostatic interactions: the
finite slope of the LO mode at ${\bf \Gamma}$ is lost, and although
quadrupolar interactions persist in the BN-metal surface system, the 
dipolar interactions are screened out.
\begin{figure}[!]
\begin{minipage}[t]{8.0cm}
\begin{center}
\leavevmode
\epsfxsize=8.5cm
\begin{turn}{-90}
\epsffile{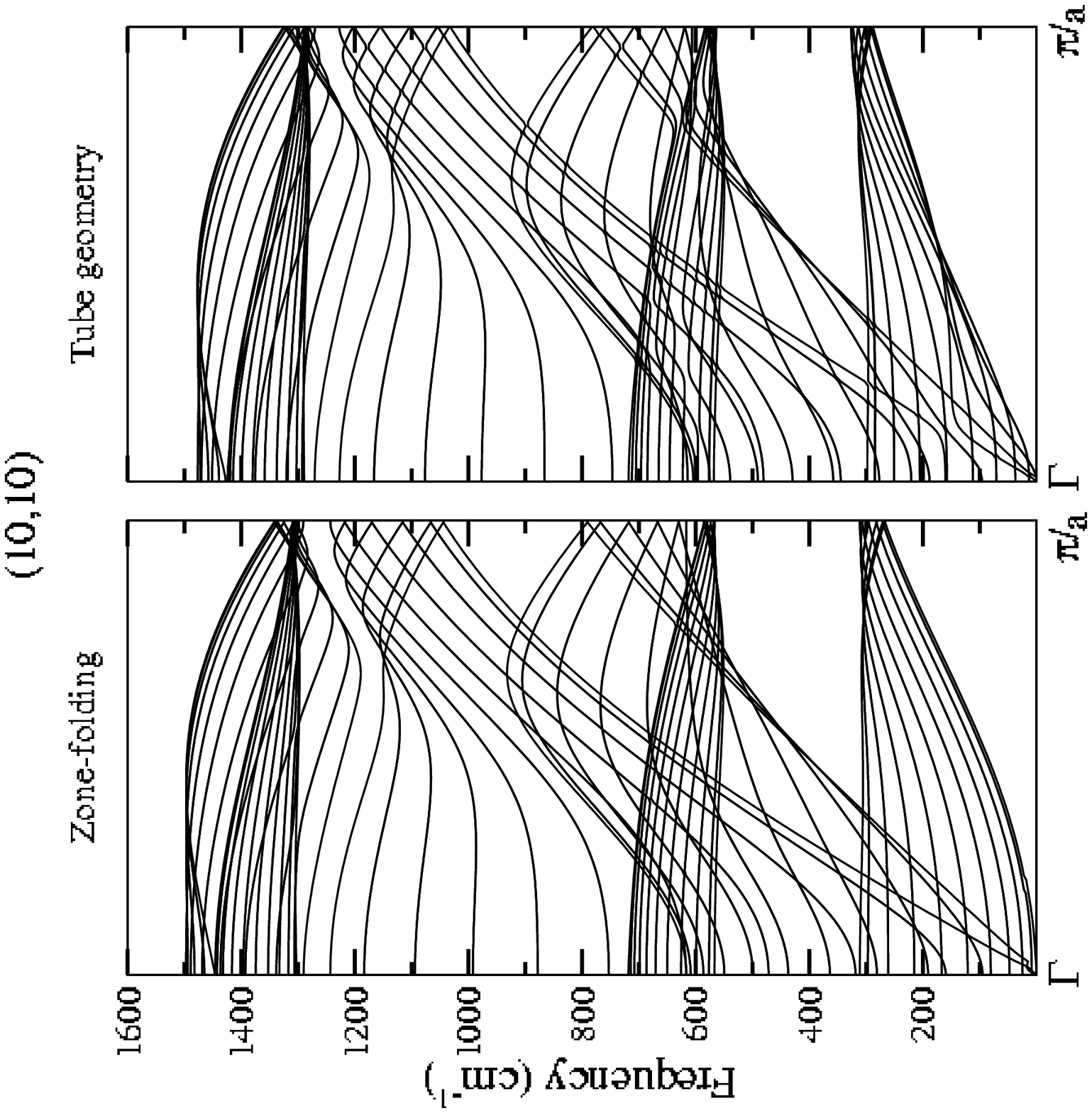}
\end{turn}
\end{center}
\end{minipage}
\caption{Comparison of the
the phonon dispersion curves of the (10,10) tube (40 atoms in the
unit cell and a diameter of 13.9~\AA) calculated
directly from the tube structure, and those obtained using
the zone-folding approach from the results for the
flat sheet shown in Fig.(\ref{fig:sheet}a).}
\label{fig:1010}
\end{figure}

Interestingly, Rokuta {\em et al.\/}~\cite{rokuta} report a degeneracy of the
ZO and ZA modes at {\bf K}. This is unexpected, because such modes should
display a splitting related to the different masses of B and N. The fact 
that no splitting is observed in the experiments is probably a result of 
a deviation from planarity in the BN monolayer due to the presence of the
metal surface. 
Another important point might be  
the quite different positions occupied by B and N 
atoms on the surface. 
In fact, Rokuta and coworkers reported a buckling of the BN monolayers on Ni(111),
where B atoms, which occupy fcc sites, lie 0.2~\AA\ below the N atoms, which 
site directly on top of the Ni atoms in the surface. 

\subsection{Nanotubes}
\label{sub:tubes}

Let us now move on to discuss the results obtained for BN nanotubes.
We have considered a series of (n,n) (arm-chair), (n,0) (zig-zag) and 
(n,m) (n~$\neq$~m~$\neq$~0, chiral) nanotubes. For (n,0) nanotubes we
have included all tubes with n from 6 to 16, {\em i.e.\/} diameters
from 5 to 14~\AA, approximately; while for (n,n) tubes we considered
specifically the (6,6), (10,10) and (15,15) tubes, with diameters
between 8 and 21~\AA. Only two chiral nanotubes were considered, since
these structures usually have very large unit-cells, namely the (10,5) and
(10,7) tubes, with diameters of 10.6 and 11.9~\AA, respectively.

\begin{figure}[]
\begin{minipage}[t]{8.0cm}
\begin{center}
\leavevmode
\epsfxsize=8.5cm
\begin{turn}{-90}
\epsffile{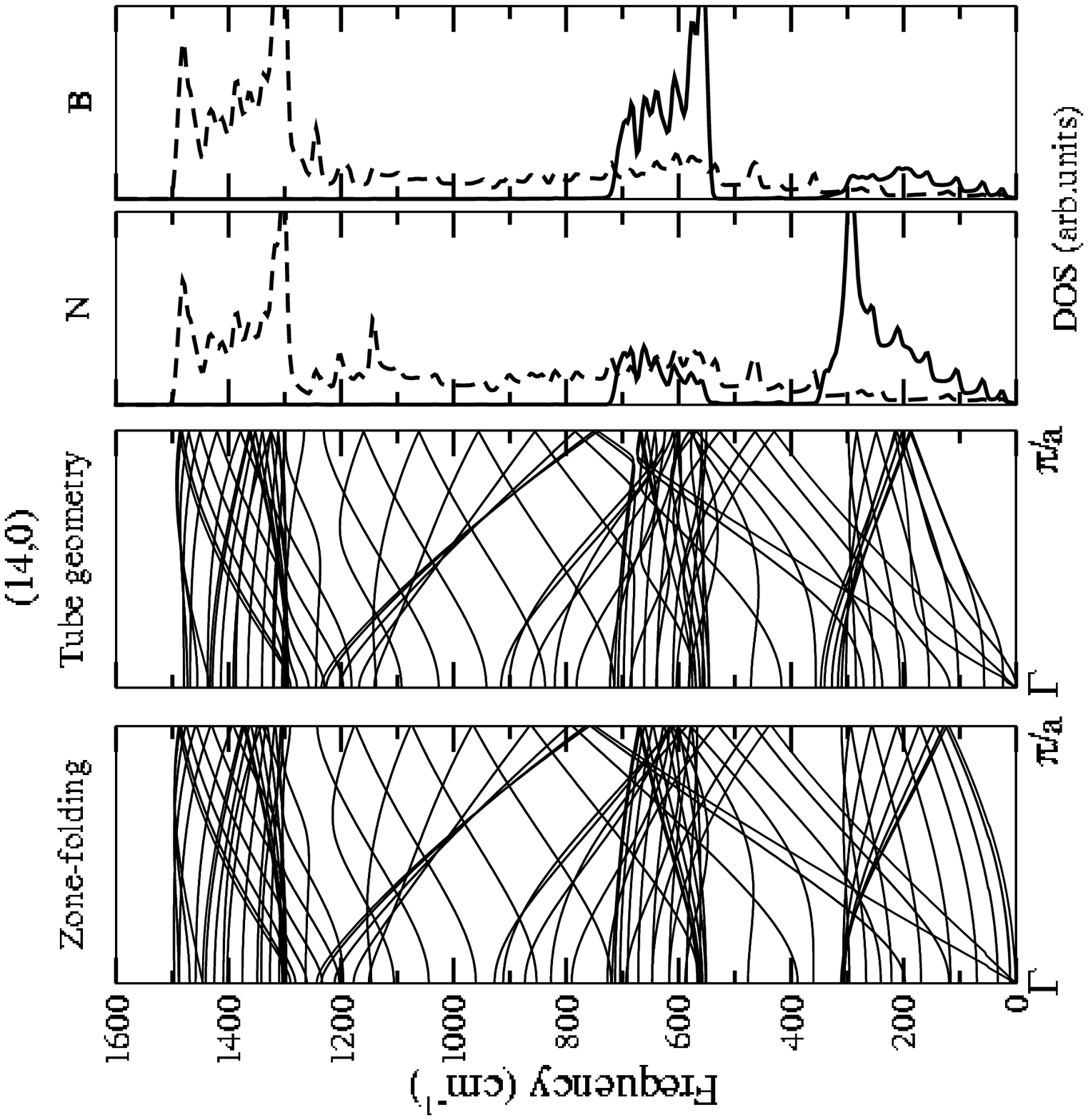}
\end{turn}
\end{center}
\end{minipage}
\caption{
Same as Fig.~\ref{fig:1010} for the (14,0) tube (56 atoms
in the unit cell and a diameter of 11.3~\AA).
The vibrational density
of states decomposed in the different directions of the atomic displacements
is also shown. Solid lines for out-of-plane displacements and dashed lines
for in-plane displacements.}
\label{fig:140}
\end{figure}

In Fig.~\ref{fig:1010}, for the (10,10) tube, and Fig.~\ref{fig:140}, for 
the (14,0), 
we compare the phonon dispersion curves calculated
directly from the tube structure, and those obtained for the same
tubes employing the zone-folding approach from the phonon bands of the
flat sheet discussed above. Firstly, let us remark that the zone-folding
method cannot reproduce certain low energy modes; in particular, it cannot
give rise to a breathing mode, which in the flat sheet corresponds to a zero
energy translation perpendicular to the plane. However, the high frequency
modes are expected (and indeed are) to be reasonably well described. In spite
of the limitation regarding low frequency modes, the zone-folding approach
has been extensively used for the interpretation of experimental results
of carbon nanotube vibrational modes~\cite{dresselhaus:eklund,kasuya}, 
partly due to its simplicity, and also
to the fact that the phonon modes of graphite can be determined with very
high accuracy. 
One of the aims of the present paper is to study if the zone-folding
approximation can also be useful in the case of the BN nanotubes.

Generally speaking, for experimentally relevant tube diameters, the 
zone-folding method reproduces quite well the overall phonon band structure.
Deviations can be larger for the smaller nanotubes, where the effects of 
curvature are more noticeable. In particular, we shall
see below how curvature effects influence the breathing mode. Another 
difference that can be appreciated between the tube dispersion bands and
those determined from zone-folding is a general 
softening of frequencies for the explicit calculation, 
where curvature effects are taken into account. 
The downward shift is especially clear at high frequencies, but is also
noticeable at intermediate ones. This shift is not homogeneous, and it 
has also a non-trivial dependence on the wave vector and on the nature
of the mode (radial or tangential).

Also shown in Fig.~\ref{fig:140} is the vibrational DOS for the (14,0) 
tube decomposed in the different directions of the atomic displacements. 
This curve resembles closely the result obtained for the BN monolayer. 
There is a clear energy separation
between radial and tangential modes (parallel or perpendicular to the
tube axis), the radial modes corresponding to frequencies below 
$\sim$~700~$\mbox{cm}^{-1}$. This corresponds to the already mentioned 
separation between
out-of-plane and in-plane-modes in the case of the planar sheet, as could
be expected, since the radial modes are roughly derived from the 
out-of-plane vibrations of the plane.

There are four acoustic bands for the tubes.
They correspond to a longitudinal
mode (atomic displacements parallel to the tube axis),
two degenerate transversal modes (atomic displacements
perpendicular to the tube axis),
and
a torsion of the tube around its axis
(the so-called {\it twiston} band which, in the limit
of ${\bf k}$=0,
generates
a rigid rotation of the tube around its axis).
This last mode is characteristic of a one-dimensional system like
the nanotubes, and no
analog can be found for the bulk or the infinite planar sheet.
According to Saito {\it et al.}~\cite{saito}, and in contrast to the
case of the flat sheet, all four bands would be required
to approach ${\bf \Gamma}$ linearly.
This relates to the fact that
none of the acoustic bands
of the nanotubes is solely derived from the quadratic out-of-plane
acoustic band of the sheet:
the transversal
acoustic bands can be regarded as a
combination of the TA in-plane and ZA
out-of-plane modes of the planar sheet.
However, it is interesting to point out here that the low energy
behavior of the transveral acoustic bands in the nanotubes
is still the subject of some controversy and some
authors have recently proposed that these bands should exhibit a
quadratic behaviour
for small values of ${\bf k}$~\cite{mahan,popov}.
The situation is more clear for the other acoustic bands
since it is possible to establish a
one-to-one correspondence between
the longitudinal and twiston
bands of the tubes and
the LA and TA bands of the
planar sheet, respectively.
In fact, the
sound velocities of the latter
bands are, within the precision of the calculations ($\sim$1~Km/s),
independent of the tube radius and chirality, and almost identical
(although somewhat smaller) to those of the corresponding
planar-sheet bands. Similar results have been
obtained for the case of the carbon nanotubes~\cite{saito,yu}.
This insensitivity
to the tube structure and radius
confirms the predictions of a simple
continuum elasticity model of the tubes~\cite{yakobson} where the
elastic constants are directly taken from the planar sheet, from which
one obtains
$v^{tube}_{LA}=v^{2D}_{LA}$ and $v^{tube}_{twiston}=v^{2D}_{TA}$.
The independence of the mechanical properties on the chirality
of the tubes is here a simple consequence of the isotropy of the underlying
hexagonal structure of the BN planes. We have
used this model to estimate the sound velocity of the transversal band,
obtaining $v^{tube}_{TA}=\sqrt{1\over2}\;v^{2D}_{TA}$=9~Km/s.
In the TB calculations, however, the
transversal bands seem to be more sensitive to the numerical
uncertainties than the other acoustic bands, being difficult to
extract an accurate slope.
In fact, although the transversal bands
apparently exhibit
a linear dependence with the phonon wavevector (see figures \ref{fig:1010}
and \ref{fig:140}), for very small
values of ${\bf k}$, and consequently for very small frequecencies,
they seem to bend slightly. This
would be roughly consistent with the proposed~\cite{mahan,popov}
quadratic behavior, although
problems of numerical accuracy cannot be ruled out. The clarification of this
point will be the subject of future work.
Here we have decided to make an estimation of the sound velocity of the 
transversal modes
from the slope of these bands in the region where they still exhibit a
clear linear behaviour. In this way 
we have found values ranging from 7~Km/s, for the narrower tubes,
to 8~Km/s, for those with larger diameters. This is again in reasonable
agreement with the result that we obtained from continuum elasticity theory.

\begin{figure}[!]
\begin{minipage}[t]{8.0cm}
\begin{center}
\leavevmode
\epsfxsize=8.5cm
\begin{turn}{0}
\epsffile{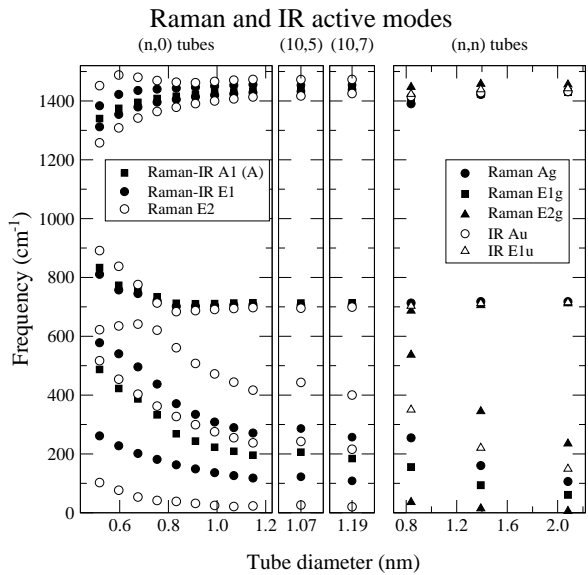}
\end{turn}
\end{center}
\end{minipage}
\caption{
Active modes for the (n,0), chiral and (n,n) BN nanotubes
considered in this work. Symbols have the same meaning for (n,0) and
chiral tubes, except that for chiral tubes A1 has to be read as A (see the
text). See Ref.~\onlinecite{tables} for the numerical values of the frequencies.
}
\label{fig:active-modes}
\end{figure}

In Fig.~\ref{fig:active-modes} we have plotted the frequencies of the 
optically active modes for the different tubes considered as a 
function of the tube radius~\cite{tables}. 
The values plotted have been obtained including
the long range electrostatic interactions, but the picture does not 
change significantly if they are not included~\cite{longrangeontubes}. 
We have classified the modes according to the corresponding symmetry
point group~\cite{symmetry}, which for zig-zag (n,0) nanotubes is
$C_{2nv}$, for arm-chair (n,n) nanotubes is $C_{2nh}$, and for 
chiral (n,m) nanotubes is $C_N$, where 
N=2(n$^2$+m$^2$+nm)/d$_R$ with d$_R$ being the greatest common 
divisor of 2n+m and 2m+n~\cite{symmetry}, {\it i.e.} 
N=70 for the (10,5) nanotube, and
N=146 for the (10,7) nanotube.
The total number of Raman and IR active modes
and their distribution over the frequency spectrum is very similar for 
both zig-zag and chiral nanotubes. 
In the
(n,0) nanotubes the rotationally
invariant modes can be classified according to how they transform under
reflection through the symmetry plane $\sigma_v$, namely $\mbox{A}_1$ and
$\mbox{A}_2$ modes; only the first are Raman and IR active, the $\mbox{A}_2$
being inactive. The chiral tubes do not contain a $\sigma_v$ mirror plane,
and therefore all six A modes are active. In total, the (n,0) nanotubes
have eight modes with non-zero frequency 
which are both IR and Raman active (three 
belong to the $\mbox{A}_1$ representation, and five to the 
$\mbox{E}_1$), 
and 
six modes ($\mbox{E}_2$) which are only Raman active, while for the chiral
tubes the count is nine IR and Raman modes (four A and five $\mbox{E}_1$), 
plus six ($\mbox{E}_2$)
Raman active modes~\cite{symmetry}. 
In the case of the (n,n) nanotubes, due to the 
existence of inversion symmetry in the point group 
the active modes can only be either IR or
Raman active, having a total of four IR active (one $\mbox{A}_u$ 
and three $\mbox{E}_{1u}$) and nine Raman (three $\mbox{A}_g$,
two $\mbox{E}_{1g}$ and four $\mbox{E}_{2g}$) active modes.
Comparing with carbon nanotubes, the bigger difference appears in the 
case of the (n,0) nanotubes, where the number of active modes is nearly 
double for BN nanotubes than for carbon nanotubes~\cite{CNTsymmetry}.
\begin{figure}[!]
\epsfxsize=8.5cm
\begin{turn}{-90}
\epsffile{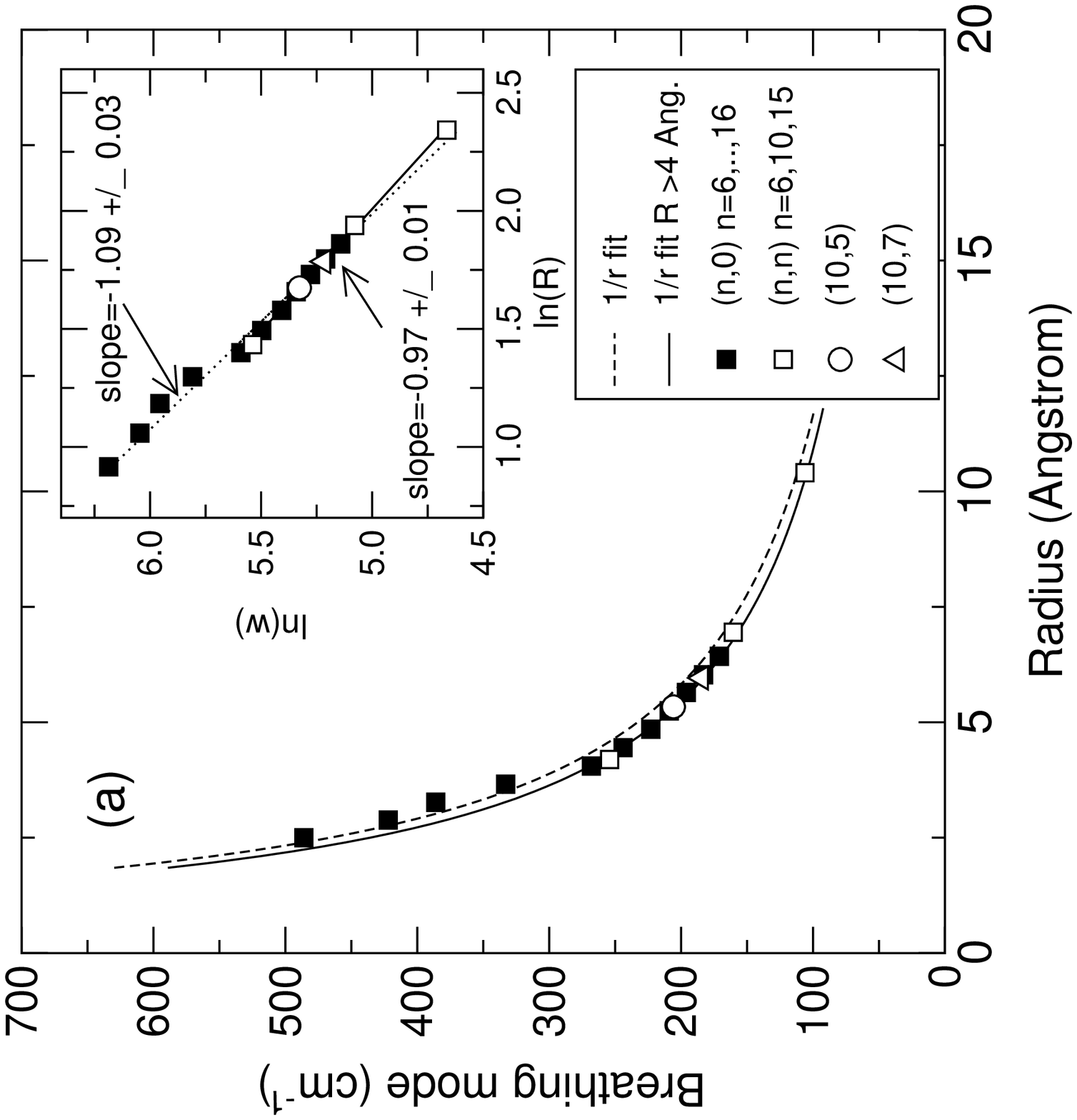}
\end{turn}
\begin{turn}{-90}
\epsfxsize=8.5cm
\epsffile{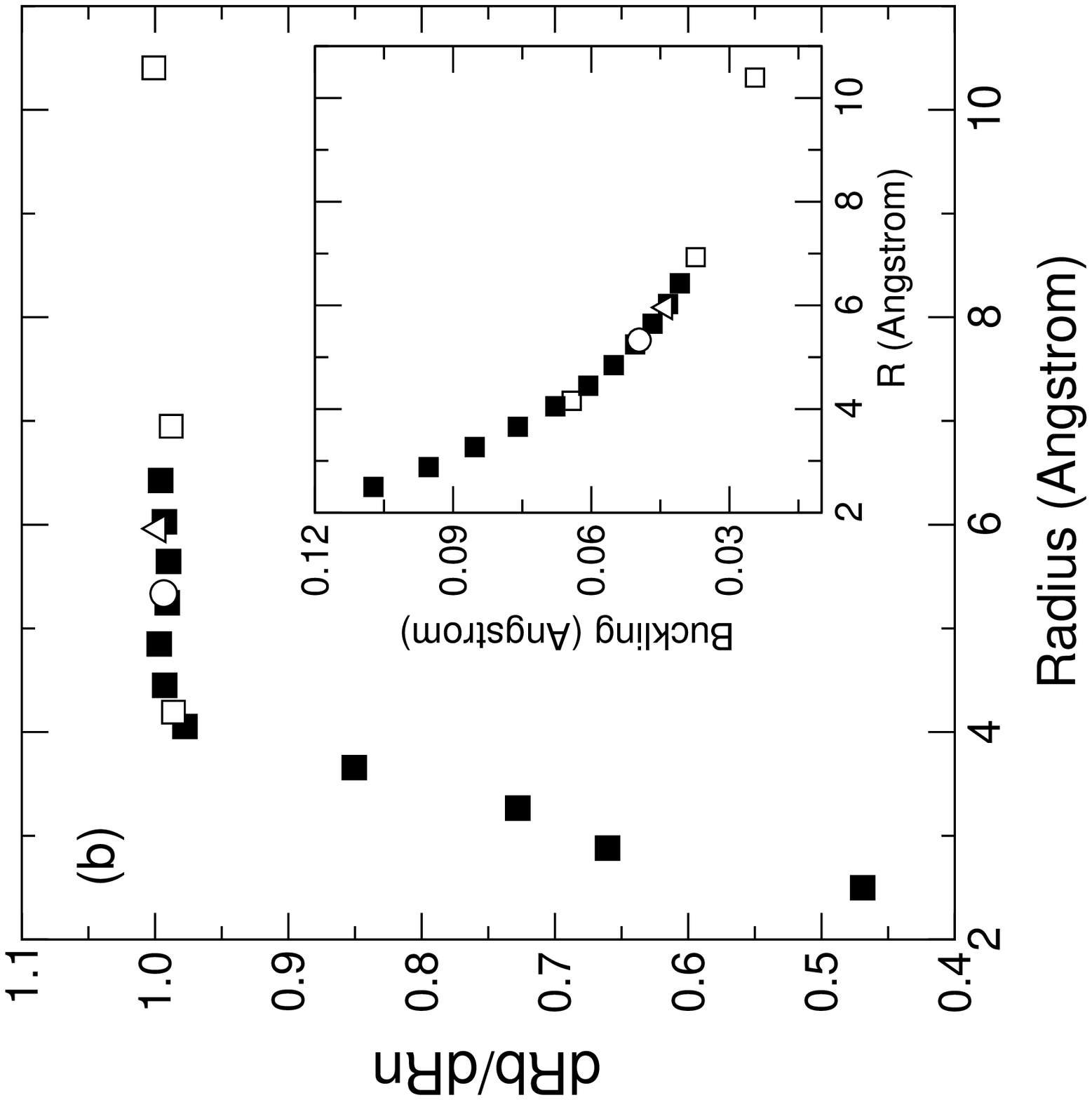}
\end{turn}
\caption{(a) Frequency of the radial breathing mode of BN nanotubes as a
function of tube radius. The inset presents a log-log plot which clearly
shows the r$^{-1}$ behavior. Panel(b) shows the relative radial
displacement of the B atoms and the
N atoms in the radial breathing mode.
The evolution of the buckling in the
nanotube structure is plotted in the inset
as a function of the radius. The meaning of the symbols is identical in all
the panels and plots.
See Ref.~\onlinecite{tables} for the numerical values of the frequencies.}
\label{fig:breathing-mode}
\end{figure}

In Fig.~\ref{fig:breathing-mode}(a) we have plotted the frequency of the 
breathing mode as a function of the tube radius. This mode is of 
importance in carbon nanotubes because it can be correlated with 
the tube radius 
and it is Raman active for all tube chiralities. 
The same is true here. Although there are minor deviations
at small radii (below 4~\AA), the breathing mode frequency follows 
very closely a dependence on the inverse of the tube radius:
$\nu_{BM}(r) = A/r$. This is corroborated in the log-log plot shown 
in the inset 
of Fig.~\ref{fig:breathing-mode}(a), which fits very well to a linear 
behavior with a slope -0.97~$\pm$0.01 (-1.09~$\pm$0.03 including the
tubes with $r < 4$~\AA). Leaving out the tubes with $r < 4$~\AA, 
a fit to this expression gives $A = 1091~\mbox{cm}^{-1}\mbox{\AA}$, 
with a standard deviation smaller than 3~$\mbox{cm}^{-1}$.  
This value is quite
insensitive to the tube structure, as occurs also with 
carbon nanotubes~\cite{sanchez-portal,kurti}. 
As can be seen in Fig.~\ref{fig:breathing-mode}(a), the breathing mode 
exhibits a hardening for 
the smaller structures ($r < 4$~\AA) respect to what would be expected from
the $r^{-1}$ fit to the frequencies of the tubes with the larger diameters.
If we include the frequencies of the smaller tubes, the quality of the
fit decreases (mean deviation of 16~$\mbox{cm}^{-1}$) and
the value obtained for $A$ increases to 1160~$\mbox{cm}^{-1}\mbox{\AA}$, in
accordance with the mentioned hardening.
Interestingly, this behavior is the oppositte to what was observed for the
carbon tubes~\cite{sanchez-portal}, where there is a softening of 
the breathing mode for increasing curvature. Therefore, this effect
must be related to an unique characteristic of the BN tubes as compare to the
carbon nanotubes:
the buckled structure of its surface. 
For small and moderate radius BN nanotubes, the B atoms lie somewhat
closer to the nanotube axis than the N atoms. This buckling is of the order
of 0.1~\AA\ for the smallest nanotubes, but decreases rapidly with 
increasing radius~\cite{elastic_tubes}.  
Fig.~\ref{fig:breathing-mode}(b) shows how the nanotube atoms 
are displaced in the breathing mode. The values plotted are the magnitude of
the radial displacement of B atoms ($\delta R_{B}$) divided by that of the N 
atoms ($\delta R_{N}$). For $\delta R_{B}/\delta R_{N} = 1$ the displacements
are of equal magnitude, and we would have a pure breathing mode, just as
in a carbon nanotube. The observed behavior is slightly different:
for tubes with $r < 4$~\AA\ the displacement of the N (outer) atoms is
larger than that of the B (inner) atoms. This can be explained in a simple
model as a consequence of the tendency 
to preserve in the pattern of displacements
of this vibrational mode the bond angles of
the buckled surface.

\begin{figure}[!]
\epsfxsize=8.5cm
\begin{turn}{-90}
\epsffile{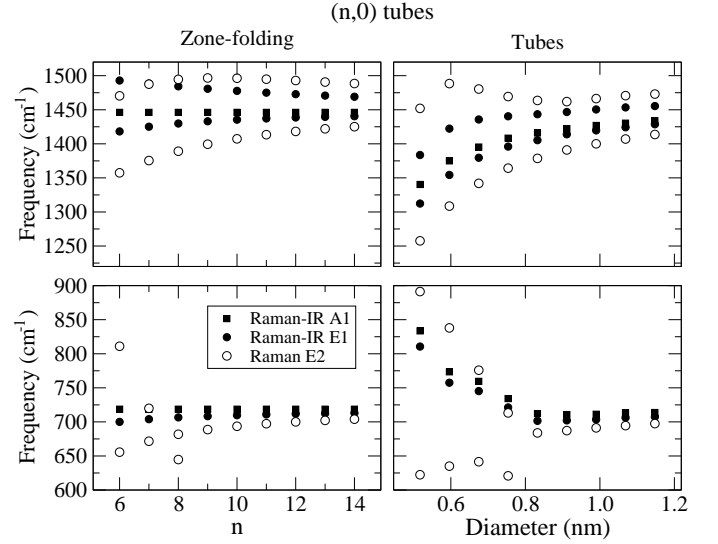}
\end{turn}
\caption{Comparison between the optically active
modes predicted by the zone-folding approximation
and those obtained directly from the tubes geometry. Here we show the
results for (n,0) BN tubes
with n running from 6 to 14.
The modes have been classified
according to their symmetry (the meaning of the symbols is the same
in the four panels).
For clarity,
intermediate frequencies (around 700~$\mbox{cm}^{-1}$) and
high frequencies (above 1250~$\mbox{cm}^{-1}$) are shown separately.
See Ref.~\onlinecite{tables} for the numerical values of the frequencies.}
\label{fig:ZF_n_0}
\end{figure}

\begin{figure}[]
\epsfxsize=8.5cm
\begin{turn}{-90}
\epsffile{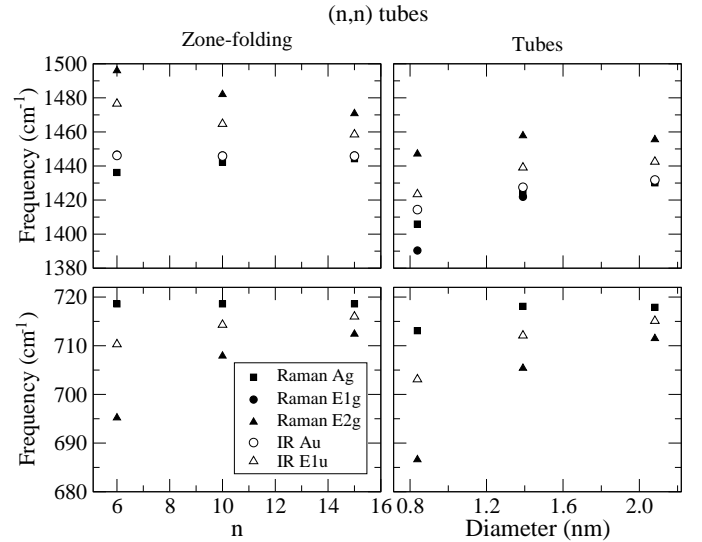}
\end{turn}
\caption{Same as Fig.~\ref{fig:ZF_n_0} but for (n,n) tubes, with n=6,10 and 15.
See Ref.~\onlinecite{tables} for the numerical values of the frequencies.}
\label{fig:ZF_n_n}
\end{figure}

As is well known~\cite{kudin,kurti,mahan}, 
the constant $A$ can be estimated from the elastic constants of the plane, 
\begin{eqnarray}
A\approx {1\over 2\pi}\sqrt{C_{xxxx}\over \rho}= {1\over 2\pi} v^{2D}_{LA}. 
\end{eqnarray}
Using this equation we obtain the value $A$=1135~$\mbox{cm}^{-1}\mbox{\AA}$, 
which is in reasonable agreement with the value found by the direct calculation.

Finally, let us discuss the ability of the zone-folding approach to predict
the frequencies of the active modes of BN nanotubes. 
In the figures \ref{fig:ZF_n_0} and \ref{fig:ZF_n_n} 
we plot the frequencies obtained from zone-folding 
side-by-side with those obtained directly from the nanotubes, in both
cases including the electrostatic interactions. For clarity, this is done
separately for intermediate frequencies (around 700~$\mbox{cm}^{-1}$) and
for high frequencies (above 1250~$\mbox{cm}^{-1}$), and separately
for $(n,0)$ and $(n,n)$ nanotubes. Overall, we can say that zone-folding
is capable of rather accurate estimations of the actual frequencies,
particularly for nanotubes with radii larger than 4~\AA. In this case
even the ordering of the modes is accurately reproduced, both for $(n,0)$ 
and $(n,n)$ tubes. The intermediate frequency modes are very well reproduced,
but there is a slight over-estimation of the high-frequency modes,
of the order of a few tens of wavenumbers.

That the zone-folding approach is capable of providing accurate estimates
of the BN nanotube phonon frequencies rests on the absence of a 
${\bf k}=0$ (macroscopic) dipolar field in the BN monolayer. The reduced
dimensionality of the system prevents the build up of a ${\bf k}=0$
component of the electrostatic field, which thus only has a finite range.
Therefore, no splitting of modes is observed at ${\bf \Gamma}$; the only
effect of the electrostatic interactions is to shift slightly the high
frequency modes and to confer a finite slope to the LO mode 
in the ${\bf k} \rightarrow {\bf \Gamma}$ limit, modifications which 
do not prevent zone-folding from providing accurate estimates of the
phonon modes of isolated nanotubes. 
A subtle point here is that the applicability of 
zone-folding in the case of BN also depends on the degree of 
transferability of the dielectric constant $\epsilon$ and 
the Born effective charges [see Eq.(\ref{eq:longrange})]
between the monolayer
and nanotube geometries. For small nanotube radii there may be noticeable
changes in the dielectric constant, but this is unlikely to affect
the application of zone-folding to all but the narrowest nanotubes.

\section{Summary and Conclusions}
\label{sec:concs}

In this work we have presented a detailed analysis of the vibrational
properties of a BN monolayer, and of single-walled BN nanotubes of different
diameters, including (n,0), (n,n) and two chiral nanotubes. Our calculations
are based on a non-orthogonal TB model complemented with long-range 
electrostatic interactions, not included in the TB model. The incorporation
of these electrostatic interactions corrects to some extent the deficiencies 
of the TB model at intermediate and high frequencies in the BN layer. Using a 
simple model for a metal surface, we have analyzed the influence of 
placing a BN monolayer over such a surface on the phonon bands of the 
monolayer. The effect appears to be small and confined only to the high 
frequency range, above 1250~$\mbox{cm}^{-1}$. Generally, our 
calculated phonon frequencies, sound velocities and elastic constants
are in good agreement with previous experimental and theoretical
results. Concerning the nanotubes, we have obtained phonon band 
structures and have analyzed the optically active modes (IR and Raman)
as a function of the nanotube structure and radius. We have also 
performed a detailed comparison between the predictions of the 
zone-folding approach and the actual frequencies of optically active 
modes obtained directly from the nanotubes, finding rather good agreement
between the two for all except the narrower tubes, having radii below
4~\AA. High frequency mode frequencies are systematically overestimated
by about a few tens of wavenumbers, but intermediate mode frequencies 
are given with high accuracy. Therefore, zone-folding should also be
useful in the context of BN nanotubes, as it has been in spectroscopic
studies of carbon nanotubes. We also report on the behavior of the frequency
and the pattern of atomic displacements of the radial breathing mode
as a function of the tube radius.

\section*{Acknowledgments} 
The authors want to thank Prof. Pablo Ordej\'on for
stimulating discussions. This work was supported in part by the 
Spanish Ministerio de Ciencia y Tecnolog\'{\i}a (grants MAT2001-0946, 
BFM2000-1312-C02 and BFM2002-03278), and by the
Basque Departamento de Educaci\'on, Universidades e Investigaci\'on and
the University of the Basque Country UPV/EHU 
(grant 9/UPV 00206.215-13639/2001). D.S.P 
acknowledges support from the Spanish Ministerio de Ciencia y Tecnolog\'{\i}a
and CSIC under the "Ram\'on y Cajal" program.  

\appendix*

\section{Absence of LO-TO splitting for a BN monolayer}
\label{nosplitting}

In this appendix
we briefly discuss the role of long-range electrostatic
interactions in a monolayer, comparing the results with those found
for bulk systems.
Let us consider a thin slab of polar material parallel to the $xy$ plane.
We focus first in
the in-plane optical modes, for which  the atomic displacements
${\bf u}$ are parallel to the
$xy$ plane. Associated with this phonon there is an electric
polarization field ${\bf P}={1\over A}(Z{\bf u})f(z)\cos({\bf k \cdot r})$
and a charge density $\rho_P=-{\bf \nabla \cdot P}$,
where $Z$ is the Born effective charge tensor within the unit cell, $A$
is the area of the unit cell, and
$f(z)$ is a function
related to the profile (number of layers)
of the slab which can be approximated by
$\delta(z)$ for a monolayer.
We stress that ${\bf k}$, ${\bf u}$, and ${\bf P}$ are all parallel
to the slab for pure in-plane modes.
For the particular case of
a system with hexagonal symmetry, like our BN sheets, the
in-plane properties are isotropic, and $Z$ can be regarded as a scalar.
Although this is not crucial for our arguments, we will make 
this simplification in the following.  The charge density $\rho_{_P}$ is 
proportional
to the product ${\bf k \cdot u} \; $, and no macroscopic charge
accumulations appear due to the TO  modes.
However,
a macroscopic electric field ${\bf E}_{mac}$ can appear associated with the
LO modes. The correct limit when ${\bf k} \rightarrow 0$
is then calculated by
solving Poisson's equation for $\rho_{_P}$. The components of the
electric field in the directions parallel  to the slab are
\begin{eqnarray}
E^{i}_{mac} = {\frac {4 \pi \; Z } {A \; \epsilon} } \:
{\frac { ({\bf k \cdot u})} { k }} \;   k^{i}
B(z,k) \cos({\bf k \cdot r}),
\end{eqnarray}
where $k$ is the modulus of ${\bf k}$, and $B(z,k)$ is given by
\begin{eqnarray}
B(z,k) = \int_{-\infty}^{\infty} d\alpha \;
{\frac
{ \bar f(\alpha k) e^{-i (\alpha k z)} } {1+\alpha^2 }},
\end{eqnarray}
$ \bar f(q)$ being the Fourier transform of $f(z)$.
Once the electric field has been calculated, the
corresponding modification of the force-constants matrix can be
obtained taken into account that the force acting on an atom is
given by ${\bf F}=Z \; {\bf E}_{mac}$~\cite{Gonze} .
In the case of the
monolayer we take $f(z)=\delta(z)$, then $B(z=0,k)=1/2$ and the
correction of the force-contants matrix for the in-plane modes becomes
\begin{eqnarray}
C_{ij} = {\frac {2 \pi \; Z^2 } {A \; \epsilon} } \:
{\frac { k^i \: k^j  } { k }}.
\end{eqnarray}
This correction goes to zero when ${\bf k} \rightarrow 0$ irrespective
of the polarization of the mode and, therefore, both LO and TO
in-plane modes have the same frequency. This is in 
agreement with the results obtained for slabs of ionic
materials using different models~\cite{fuchs,lucas}: pure in-plane
optical modes have frequencies  close to those of the TO phonons
in the bulk. It can also be noticed the linear behavior of the LO 
mode as it approaches $\Gamma$. We move now to analyze the bulk limit. 
By taking
$f(z)=1/L$, being $L$ the distance between atomic layers, we
can study the limit for bulk phonons with $k_z=0$ 
($k_z$ is the wavevector perpendicular to the atomic layers in the bulk).
$B(z,k)$ becomes now $1/(Lk)$, and the term
in the force-constants matrix takes the usual form for
bulk systems~\cite{Gonze}
\begin{eqnarray}
C_{ij} = {\frac {4 \pi \; Z^2 } {AL \; \epsilon} } \:
{\frac { k^i \: k^j  } { k^2 }},
\end{eqnarray}
which has a different value for LO and TO modes.
For the intermediate case of a slab with finite thickness we find
bulk behavior for phonons with sufficiently large $k$, while the behavior 
of the monolayer is recover when $k$ goes to zero. 
The
slope of the LO modes at ${\Gamma}$ is proportional to the number
of layers in the slab. 

For modes with polarization vectors having components
along the normal to the slab (out-of-plane components
within the nomenclature used throughout this paper)
things can be quite
different. For these modes, ${\bf P}$ has a component
along $z$ and, when $k=0$, the charge distribution $\rho_{_P}$
associated to them can be rationalized as a number of
infinite planes with charges of alternating signs along $z$.  
There are also surface charges given by
${\bf P \cdot \hat n}= P_z$. All these charged planes give rise to
electric fields which do not decay along the entire
thickness of the slab. In fact, those modes with
atomic
displacements perpendicular to the slab, although are
{\it transversal} by definition (since the only phonon wavevector
that can be safely defined is parallel to the slab),
are known to have frequencies similar to those of the corresponding
longitudinal phonons
in  the bulk~\cite{fuchs,lucas}. For this reason,
Fuchs and Kliewer~\cite{fuchs} and Lucas~\cite{lucas}
arrived to the conclusion that the phonon-frequency distribution
of an ionic crystal slab converges very rapidly to that of bulk.
However, for anisotropic systems like h-BN, where the frequency
range of the in-plane and out-of-plane vibrations is very different, the
conclusion might be different. 

In the case of a free
monolayer the separation between in-plane and out-of-plane modes
is exact for an arbitrary wavevector ${\bf k}$. There is only
one out-of-plane optical mode (ZO) which, although its frequency is  
slightly modified by the 
long-range interactions, cannot exhibit any
splitting since it is a non-degenerate mode. The LO and TO in-plane 
modes
do not exhibit splitting as explained above.


\begin{thebibliography}{99}
\bibitem{tensorfortubes} 
For the construction of the Born effective charge tensor in 
the case of the tube geometry we have identified 
purely radial and tangential displacements
with, respectively, out-of-plane and in-plane movements in the planar sheet.
Therefore, for an atom $i$, which belongs to a 
tube of radius $r$ and axis (0,0,$z$), 
and occupies the position 
($r \cos(\theta)$,$r \sin(\theta)$, $z_i$), the Born effective charge tensor
takes the form:
\begin{eqnarray}
Z^i_{\mu \nu} =  \nonumber \\
\left(
\begin{array}{ccc}
(Z^i_{\perp} \cos^2(\theta)+Z^i_{\parallel} \sin^2(\theta))&
(Z^i_{\perp} -Z^i_{\parallel})\cos(\theta)\sin(\theta)& 0 \\
(Z^i_{\perp} -Z^i_{\parallel})\cos(\theta)\sin(\theta)&
(Z^i_{\perp} \sin^2(\theta)+Z^i_{\parallel} \cos^2(\theta))& 0 \\
0 & 0 & Z^i_{\parallel}
\end{array}
\right) \nonumber
\end{eqnarray}
\bibitem{largetubes} In Ref.~\onlinecite{kudin} some evidence
was presented that, while for the small and moderate radius
the strain energy is clearly smaller in the case of BN compared to
C nanotubes,
the opposite could be true for larger tubes ( $r > 30$~\AA).
However, this seems quite unlikely if, as pointed out here, a
smaller flexular rigidity of BN can already be deduced
from the behavior of the
ZA phonon band of the flat sheet.
Unfortunately, the parameter $\delta$ is quite sensitive
to the fit procedure and to numerical uncertainities, and the
values for C and BN are close enough
so no definitive conclusion can be given
here.
\bibitem{stiffness} In fact, the in-plane stiffness 
$C$ reported in Ref.~\onlinecite{kudin}
should be compared with our Young modulus $Y$ (Table~\ref{table:constants}). 
The elastic constant C$_{xxxx}$
is, in the notation of that reference, given by $C/(1-\nu^2)$ with a value
of 0.284~TPa~nm and $\nu$ being 
the Poisson ratio.
\end{thebibliography}
\end{document}